\documentclass[12pt]{spieman}  
\usepackage{amsmath,amsfonts,amssymb}
\usepackage{graphicx}
\usepackage{setspace}
\usepackage{tocloft}
\usepackage{lineno}

\title{A Statistical Method for Constraining the Capability of the Habitable Worlds Observatory to Understand Ozone Onset Time in Earth Analogs}

\author[a,*]{Sarah Blunt}
\author[b]{Eric L. Nielsen}
\author[c]{Elisabeth R. Newton}
\author[d]{Jessie Christiansen}
\author[f]{Tansu Daylan}
\author[j]{Courtney Dressing}
\author[j]{Caleb K.\ Harada}
\author[g]{Stephen R. Kane}
\author[e]{Malena Rice}
\author[i]{Romy Rodr\'iguez Mart\'inez}
\author[h]{Sabina Sagynbayeva}
\affil[a]{Department of Astronomy \& Astrophysics, University of California, Santa Cruz, CA 95064, USA}
\affil[b]{Department of Astronomy, New Mexico State University, Las Cruces, NM 88003, USA}
\affil[c]{Department of Physics and Astronomy, Dartmouth College, Hanover, NH 03755, USA}
\affil[d]{NASA Exoplanet Science Institute, IPAC, MS 100-22, Caltech, 1200 E. California Blvd, Pasadena, CA 91125}
\affil[e]{Department of Astronomy, Yale University, New Haven, CT 06511, USA}
\affil[f]{Department of Physics and McDonnell Center for the Space Sciences, Washington University, St. Louis, MO 63130, USA}
\affil[g]{Department of Earth and Planetary Sciences, University of California, Riverside, CA 92521, USA}
\affil[h]{Department of Physics and Astronomy, Stony Brook University, Stony Brook, NY 11794, USA}
\affil[i]{Center for Astrophysics \textbar \ Harvard \& Smithsonian, 60 Garden St, Cambridge, MA 02138, USA}
\affil[j]{Department of Astronomy, 501 Campbell Hall \#3411, University of California, Berkeley, CA 94720, USA}

\cftpagenumbersoff{figure}
\cftpagenumbersoff{table} 
\begin{document} 
\maketitle

\begin{abstract}
The oxygenation of Earth's atmosphere 2.3 billion years ago, which on exoplanets is expected to be most detectable via the UV ozone feature at $\sim$0.25 $\mu$m, is often regarded as a sign of the emergence of photosynthetic life. On exoplanets, we may similarly expect life to oxygenate the atmosphere, but with a characteristic distribution of emergence times. In this paper, we test our ability to recover various ``true'' emergence time distributions as a function of 1) stellar age uncertainty and 2) number of Earth analogs in the sample. The absolute uncertainties that we recover, for diverse underlying distributions, are mostly independent of the true underlying distribution parameters, and are more dependent on sample size than stellar age uncertainty. For a sample size of 30 Earth analogs, and an HWO architecture sensitive to ozone at 1\% of the current atmospheric level on Earth, we find that no ozone detections across the entire sample would place a 10$\sigma$ limit on the mean time of ozone emergence, regardless of stellar age uncertainty.
\end{abstract}


{\noindent \footnotesize\textbf{*}Sarah Blunt,  \linkable{sarah.blunt.3@gmail.com} }

\begin{spacing}{2} 

\section{Introduction}
\label{sect:intro} 

On Earth, our atmosphere has been immensely impacted by life. Oxygen became abundant in Earth’s atmosphere about 2.3 billion years ago \cite{Lyons:2014a} (also see Figure 1-8 of the LUVOIR final report\cite{LUVOIR} for a helpful visualization of this process), when the byproducts of oxygenic photosynthesis were able to accumulate in our atmosphere. There is also ongoing debate about whether the Great Oxidation Event (GOE) occurred as a step function \cite{Kopp:2005a}, or whether there was substantive oxygen present in the preceding billion years (\cite{Anbar:2007a,Crowe:2013a} but see also \cite{Slotznick:2022a}) and/or global heterogeneity \cite{Philippot:2018a}. One possible explanation is that the GOE corresponded to the rise of cyanobacteria, with previous photosynthetic organisms being anoxygenic (\cite{Kopp:2005a,Fischer:2016a}). However, oxygenic photosynthesis could also be significantly older \cite{Brocks:1999a}, with geologic processes playing the critical role in driving the GOE (\cite{Kasting:1993a,Kump:2001a}). Ultimately, the time of oxygen onset is likely set by some relationship (or lack thereof) between the rise of oxygenic photosynthesis and changing geochemical and tectonic conditions. As this interplay is not yet well understood on Earth, we do not have a robust prediction for what it would look like on exoplanets (although it has been suggested that at least some rocky exoplanets have similar geochemistry to Earth and Mars \cite{Doyle:2019a}). 

Searching for signs of life on Earth analogs is a major goal of the Habitable Worlds Observatory (HWO) \cite{astro2020}. In addition to searching for potential biosignatures in the atmospheres of individual nearby, Earth-sized exoplanets, HWO will have the potential to provide \textit{population-level} constraints. In this paper, we evaluate the ability of possible HWO architectures to answer the following statistical question: how long after it forms (if ever) does an Earth-like planet typically acquire an oxygen-rich atmosphere like modern Earth’s? Answering this question is a prerequisite for understanding whether life on exoplanets typically evolves faster or slower than it did on Earth, but has the advantage of being much more directly measurable, allowing us to perform a preliminary trade study. The main goal of this work is to enable future studies that will more robustly consider how well population-level characteristics are constrained by various HWO telescope architectures. 

Although the methodology we develop could be applied to any biosignature, we restrict ourselves here to studying the timescale of ozone emergence. The clearest determination of whether an Earth analog planet has significant atmospheric oxygen is the NUV ozone feature at $\sim$0.25 microns, where the flux in the reflectance spectrum is significantly depressed for an ozone-rich planet. \cite{Damiano:2023a} found that even with a spectral resolution of R$\sim$7 in the UV and SNR of $\sim$10, ozone features of Protozoic-like Earths (which they defined as containing 1\% of the O$_2$ abundance of modern Earth, and an O$_3$ volume mixing ratio of $\sim$1e-7, following \cite{Reinhard:2017a}) could be robustly recovered in atmospheric retrievals. 

The actual number of exo-Earths with ozone detections (and high-significance non-detections) will depend strongly on both HWO architecture and exoplanet demographics. \cite{Morgan:2024a} used a similar metric to that motivated by \cite{Damiano:2023a}, SNR$\sim$5 and R$\sim$7\footnote{Increasing SNR from 5 to 10 in turn would require an additional 4x exposure time.}
, to define characterization of exo-Earths in the NUV when investigating yields. Their yield analysis showed that, for example, 14 exo-Earths are expected to have an ozone measurement to SNR$\sim$5 precision with a nominal design for HWO (6m diameter, contrast of 10$^{-10}$, inner working angle of 70 mas, and throughput of 40\%), and assuming a SAG13\footnote{SAG13 refers to the NASA Exoplanet Program Analysis Group's Science Analysis Group-13; see \href{https://exoplanets.nasa.gov/system/presentations/files/67_Belikov_SAG13_ExoPAG16_draft_v4.pdf}{https://exoplanets.nasa.gov/system/presentations/files/67\_Belikov\_SAG13\_ExoPAG16\_draft\_v4.pdf}.} occurrence rate for Earth analogs (42\% of Sun-like stars hosting an exo-Earth). As expected, these yields increase as telescope architecture improve; the ozone levels of 25 Earth-like planets are characterized for a 9m aperture that reaches 10$^{-11}$ contrast, for example. Similarly, changes to the exo-Earth occurrence rate will directly impact the number of planets that can be characterized with any telescope architecture.  

\cite{Latouf:2024a} study a HWO architecture with \textit{visible-only} wavelength coverage, finding that O$_3$ abundances consistent with Proterozoic Earth would be \textit{undetectable} except in the case of very high SNR. Consequently, in this paper, we make the assumption that HWO will be sensitive to ozone features formed by oxygen in Earth-like atmospheres at 1\% of the present atmospheric level (PAL), the minimum level of atmospheric oxygen on Earth during the proterozoic compatible with the geologic record \cite{Kump:2008a}, which requires that HWO have UV capability.

We study the statistical ability of an HWO architecture with UV sensitivity to ozone levels consistent 1\% of the Earth's oxygen PAL to constrain the distribution of atmospheric ozone onset times on Earth-like planets as a function of two key parameters: (1) sample size of Earth analog planets (which is primarily driven by $\eta_{\oplus}$, the occurrence rate of Earths, and limiting stellar magnitude in the UV, which in turn is tied to telescope size), and (2) stellar age uncertainty. This work is complementary to previous studies of future telescopes’ ability to characterize oxygen emergence, notably \cite{Bixel:2020a} and \cite{Bixel:2021a}. \cite{Bixel:2020a} investigated the ability of an HWO-like observatory to constrain the \textit{correlation} between planet age and the presence of oxygen. \cite{Bixel:2021a} significantly expanded this analysis, incorporating the correlation analysis of \cite{Bixel:2020a} into the \texttt{Bioverse} framework, which accounts for (among other things) signal-to-noise calculation, sample simulation informed by current knowledge of exoplanet and stellar population statistics, and survey prioritization capabilities. They also investigated the ability of a HWO-like architecture to recover the characteristic timescale of oxygenation, concluding overall that HWO is unlikely to be able to robustly detect a correlation between planet age and presence of atmospheric oxygen.

In this paper, we pose a slightly different statistical question; instead of asking ``Can we detect a correlation between age and presence of oxygen?'', we ask ``Given a sample, how well can we constrain the oxygen onset time distribution?'' We do not intend for the results we derive here to be directly used to evaluate potential HWO architectures, but make our code and methodology publicly available to be incorporated into more sophisticated end-to-end predictions of HWO performance alongside existing statistical formulations, such as the strength of a stellar age-oxygen abundance correlation first proposed by \cite{Bixel:2020a}. The trade study performed here should be interpreted as an example application of our methodology, not as a final result.

The structure of this paper is as follows: in Section \ref{sec:setup}, we describe the setup of the simulations we performed in order to evaluate the trade-off between stellar age uncertainty and sample size for constraining the oxygen onset time distribution. In Section \ref{sec:results}, we present the results of these simulations. In Section \ref{sec:discuss}, we discuss what goes into the two parameters we study in tradeoff: stellar age measurements and sample size. In Section \ref{sec:conclude}, we summarize and conclude.

\section{Simulation Setup}
\label{sec:setup}

To perform a first evaluation of the tradeoff between sample size and stellar age precision for constraining the ozone onset time distribution, we performed a suite of simulations. We assumed that for a population of Earth analog planets, oxygen appears a certain amount of time after the planet forms (possibly never), and that the distribution of these times has some intrinsic scatter, described by a Gaussian with some mean $\mu_{\rm pop}$ and standard deviation $\sigma_{\rm pop}$. We created a set of simulated ``observations'' by first drawing a set of stellar ages from a uniform distribution between 0 and 13 Gyr (generally consistent with the age distribution of Milky Way stars; see \cite{Snaith:2015a,Fantin:2019a,Mor:2019a}) then assigning each true age an observational uncertainty and adding corresponding Gaussian noise. For each star in this sample, we then assigned a binary ``ozone observed'' or ``ozone not observed'' based on the probability distribution defined by $\mu_{\rm pop}$ and  $\sigma_{\rm pop}$. This resulted in a population such as the one shown in Figure \ref{fig:population} below. Given this simulated population, we then constructed a likelihood function that would allow us to ``retrieve it.'' The derivation of this likelihood is detailed in Section \ref{sec:derivation} below, which can be safely skipped.

Given this simulated population, and equipped with our likelihood function, we then ran a Markov Chain Monte Carlo (MCMC) algorithm using \texttt{emcee} \cite{emcee} to estimate posteriors on $\mu_{\rm pop}$ and $\sigma_{\rm pop}$. We used 100 walkers, a burn-in of 200 steps, and a production chain with length 500 steps. We placed uniform priors on $\mu_{\rm pop}$ and $\sigma_{\rm pop}$. Convergence was assessed by eye. An example MCMC result is shown in Figure \ref{fig:recovery}. This whole setup is summarized in Table \ref{tab:params}. All code needed to reproduce our simulations is publicly available at \url{https://github.com/sblunt/HWO\_calculations}. 

\subsection{Assumptions}

As discussed in the Introduction, even the timescale and factors leading to atmospheric oxygenation on Earth are uncertain. We chose to parameterize the true ozone onset time distribution as a Gaussian because it is a low-dimensional (i.e. 2 parameter) distribution that is capable of describing relatively diverse behavior (e.g. both wide, unconstrained distributions and also narrow, peaked distributions). In making this parameterization, we are inherently assuming that once an Earth-analog atmosphere is oxygenated, it continues to be oxygenated throughout its lifetime. 

These two assumptions (that the true ozone onset time is Gaussian, and that ozone persists after it forms) are not fundamental to the framework we have developed, and we encourage future studies to explore alternate parametrizations. Specifically, all one would need to do is replace the Gaussian cdf in Equation \ref{eq:cdfpdf} with an alternate form. 

\begin{table}[ht]
\caption{Summary of simulation inputs and outputs} 
\label{tab:params}
\begin{center}       
\begin{tabular}{|l|l|} 
\hline
\rule[-1ex]{0pt}{3.5ex}  Free parameters in simulation & Recovered parameters  \\
\hline
\rule[-1ex]{0pt}{3.5ex}   $\mu_{\rm pop}$, $\sigma_{\rm pop}$ & posteriors on $\mu_{\rm pop}$, $\sigma_{\rm pop}$ \\

\rule[-1ex]{0pt}{3.5ex}   percentage uncertainty on stellar ages &  \\
\rule[-1ex]{0pt}{3.5ex}   number of stars in sample &  \\
\hline
\end{tabular}
\end{center}
\end{table} 

\begin{figure}
\begin{center}
\begin{tabular}{c}
\includegraphics[width=\textwidth]{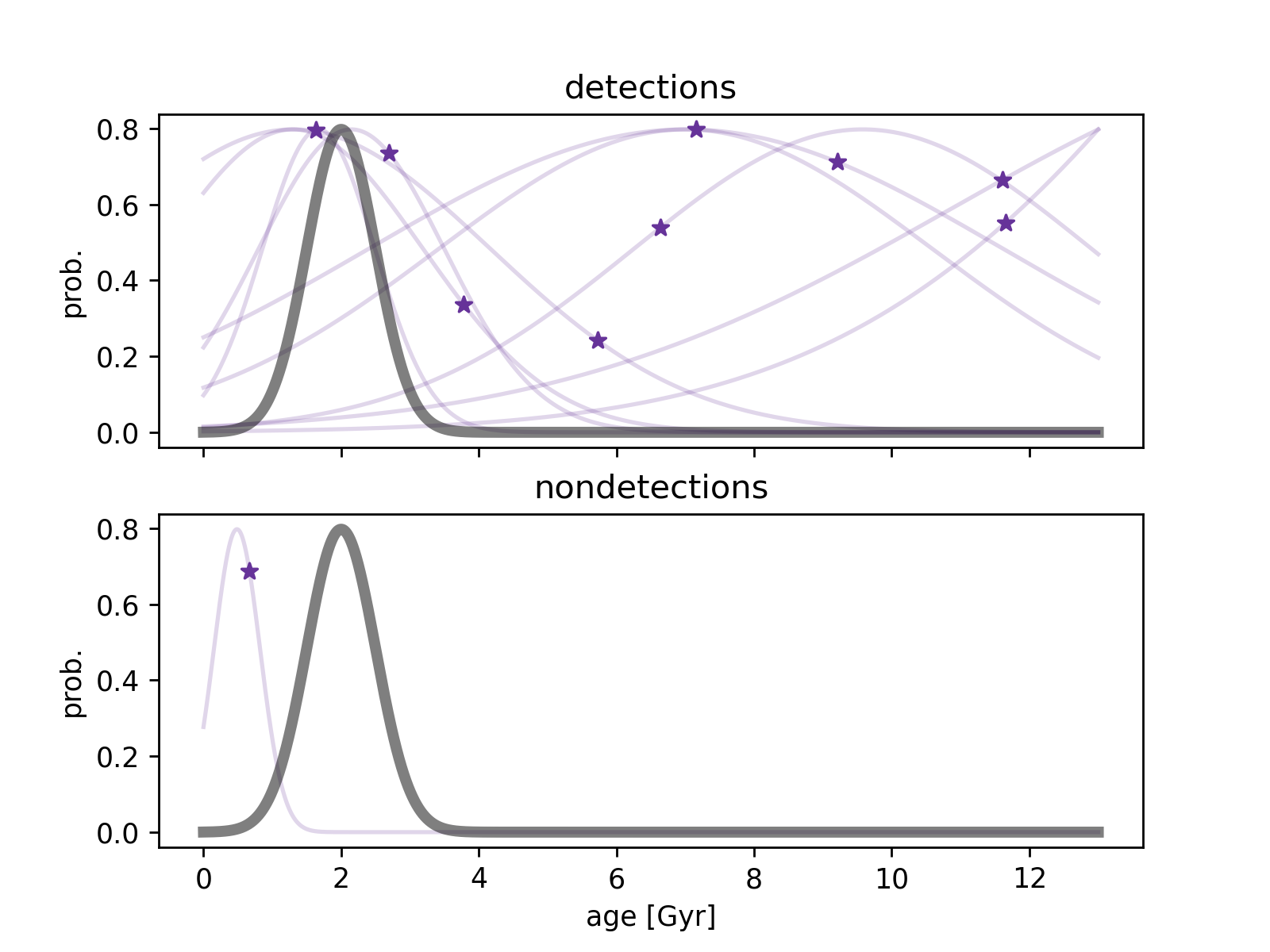}
\end{tabular}
\end{center}
\caption 
{A simulated population of Earth-like planets with ozone detections (top) and nondetections (bottom), based on the experiment setup described in Section \ref{sec:setup}. The thick gray line shows the assumed true distribution of ozone onset times, and the thin colored lines show the simulated observed stellar ages in each category. Stars show the true ages of each object, and the Gaussians represent the observed ages and uncertainties. In this specific simulation, age uncertainty was assumed to be 20\% of the true stellar age, resulting in smaller uncertainties for younger stars, and larger for older stars. \label{fig:population}} 
\end{figure} 

\begin{figure}
\begin{center}
\begin{tabular}{c}
\includegraphics[width=\textwidth]{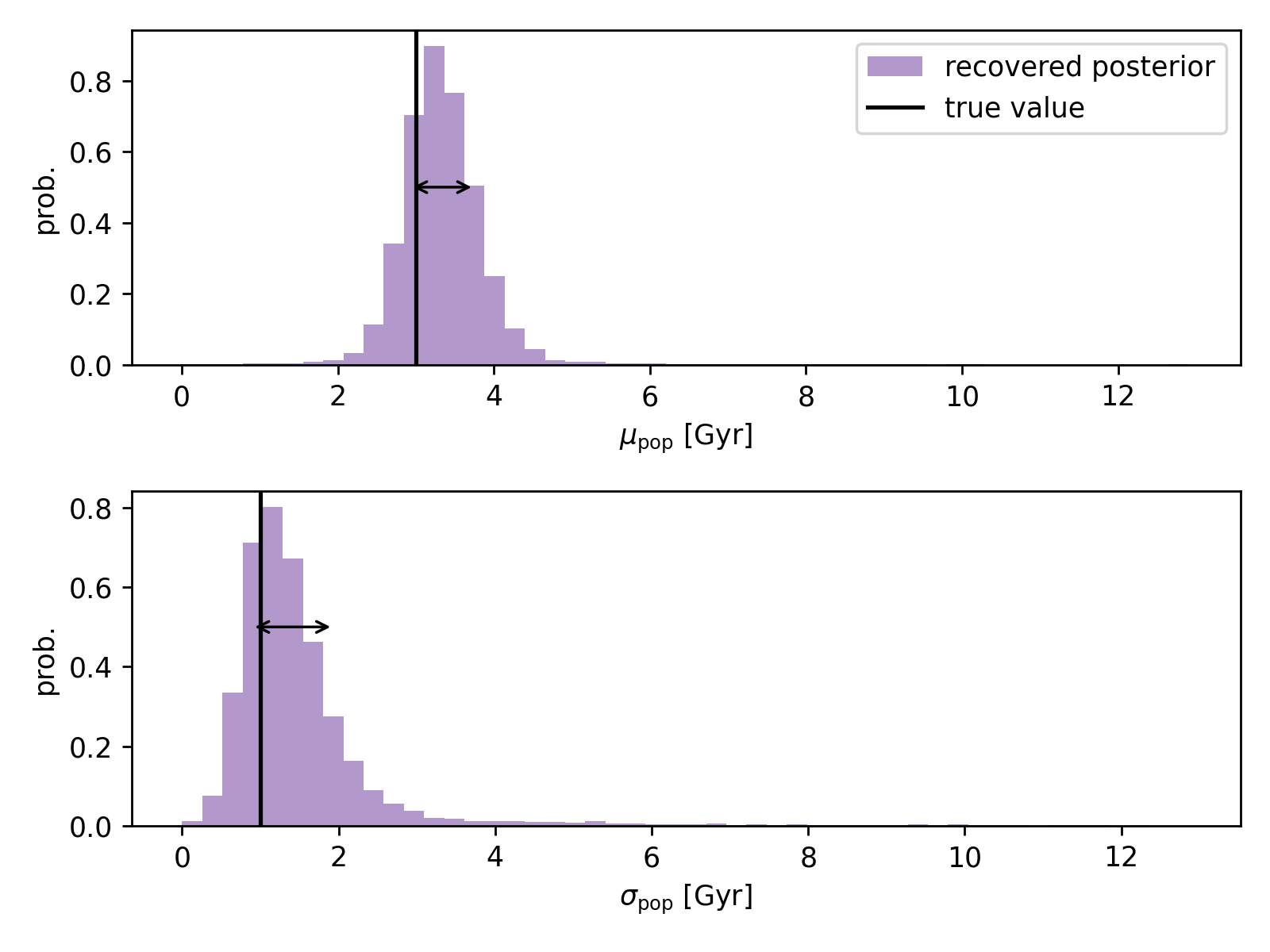}
\end{tabular}
\end{center}
\caption 
{Retrieved posteriors on $\mu_{\rm pop}$ and $\sigma_{\rm pop}$ for a simulated population of 60 planets with ozone detections or nondetections and 20\% stellar age uncertainties. Arrows show the recovered 1$\sigma$ limits on the posterior credible intervals for each parameter, which are shown in Figures \ref{fig:2gy}-\ref{fig:13gy} on the y axes. \label{fig:recovery}} 
\end{figure}

\subsection{Likelihood Derivation}
\label{sec:derivation}

We start by computing the probability of observing ozone on a given planet, represented as the probability that a random variable O$_{i}$=1, given a particular set of population-level parameters $\boldsymbol{\theta}$ := ($\mu_{\rm pop}$, $\sigma_{\rm pop}$). This probability depends not only on the values of $\boldsymbol{\theta}$, but also on the age of the star, $a_i$. Therefore, in order to compute p($O_{i}=1 | \boldsymbol{\theta}$), we can marginalize over $a_i$:

\begin{equation}
    p(O_{i}=1|\boldsymbol{\theta}) = \int p(O_{i}=1 | \boldsymbol{\theta},a_i)p(a_i)da_i.
\end{equation} $a_i$ is Gaussian distributed, with a mean value a$^{\mu}_i$ representing the measured stellar age for planet $i$ and a$^{\sigma}_i$ representing the observational uncertainty. Therefore, $p(a_i) = \phi(\frac{a_i - a^{\mu}_i }{a^{\sigma}_i})$, where $\phi$ represents the standard Gaussian pdf. $p(O_{i}=1 | \boldsymbol{\theta},a_i)$, on the other hand, is the probability of obtaining a positive ozone detection for particular values of $\boldsymbol{\theta}$, and $a_i$. This probability is $\Phi(\frac{a_i - \mu_{\rm pop} }{\sigma_{\rm pop}})$, where $\Phi$ represents the standard Gaussian cdf. This integral has an analytic solution:

\begin{equation}
\begin{split}
    p(O_{i}=1|\boldsymbol{\theta}) &=  \int \Phi(\frac{a_i - \mu_{\rm pop} }{\sigma_{\rm pop}})\phi(\frac{a_i - a^{\mu}_i }{a^{\sigma}_i}) da_i\\
    &= \;\Phi\left(\frac{a^{\mu}_i - \mu_{\rm pop}}{\sqrt{\sigma_{\rm pop}^2 + a^{\sigma \, 2}_i}}\right). \label{eq:cdfpdf}
\end{split}
\end{equation} Another way to understand this likelihood relates to the distribution over the difference of two Gaussian-distributed random variables; the stellar age, $a_i$ is a Gaussian-distributed random variable with mean $a^{\mu}_i$ and standard deviation $a^{\sigma}_i$, and the ozone onset time of planet $i$ (let's call this oz$_i$), is a Gaussian-distributed random variable with mean $\mu_{\rm pop}$ and standard deviation $\sigma_{\rm pop}$. $p(O_{i}=1|\boldsymbol{\theta})$ is the probability that  a$_i>$oz$_i$, which yields the same expression as above.

To put together the final likelihood, we next notice that p$(O_{i}=0|\boldsymbol{\theta})$, the probability of not observing ozone on planet $i$, is 1 - p$(O_{i}=1|\boldsymbol{\theta})$. Therefore, the full likelihood for a single star in the sample is:

\begin{equation}
    p(O_{i}|\boldsymbol{\theta}) = \begin{cases}
    p(O_{i}=1|\boldsymbol{\theta}), & \text{if $O_{i}=1$}\\
    1-p(O_{i}=1|\boldsymbol{\theta}), & \text{if $O_{i}=0$}.
\end{cases}
\end{equation} To obtain the likelihood for the full sample of planets with ozone non/detections, p($\{O_{i}\}|\boldsymbol{\theta})$, we simply multiply the individual planet likelihoods:

\begin{equation}
    p(\{O_{i}\}|\boldsymbol{\theta}) = \prod_i  p(O_{i}|\boldsymbol{\theta}).
\end{equation}

This likelihood is a close cousin of the more familiar binomial distribution, and reduces exactly to the binomial pdf when $\sigma_{\rm pop}$ and $a^{\sigma}_i$ are equal to 0. Unlike the binomial distribution, however, this formation allows us to test HWO's ability to recover not only the mean planetary age at which ozone occurs, but also the intrinsic physical spread of this distribution.

\section{Results}
\label{sec:results}

\begin{figure}
    \centering
    \includegraphics[width=\textwidth]{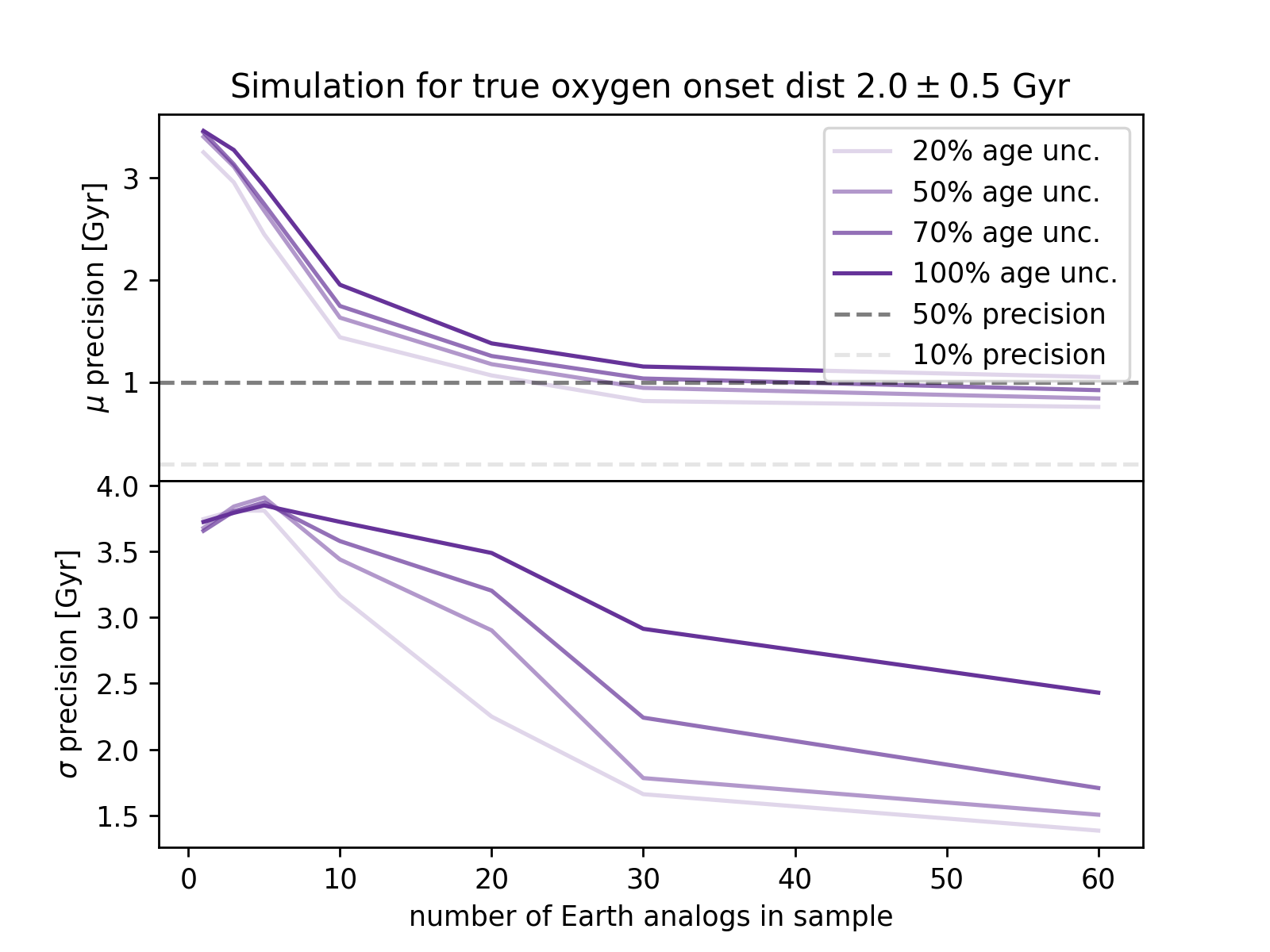}
    \caption{Posterior uncertainty (absolute precision) of the recovered parameters of the assumed underlying ozone onset times, for different precursor stellar age measurement precisions and number of Earth analogs in the sample. Takeaway: although better precursor age measurements help, increasing the sample size of Earth analogs is most impactful for better constraining the ozone onset time distribution. \label{fig:2gy}
}
\end{figure}

\begin{figure}
    \centering
    \includegraphics[width=\textwidth]{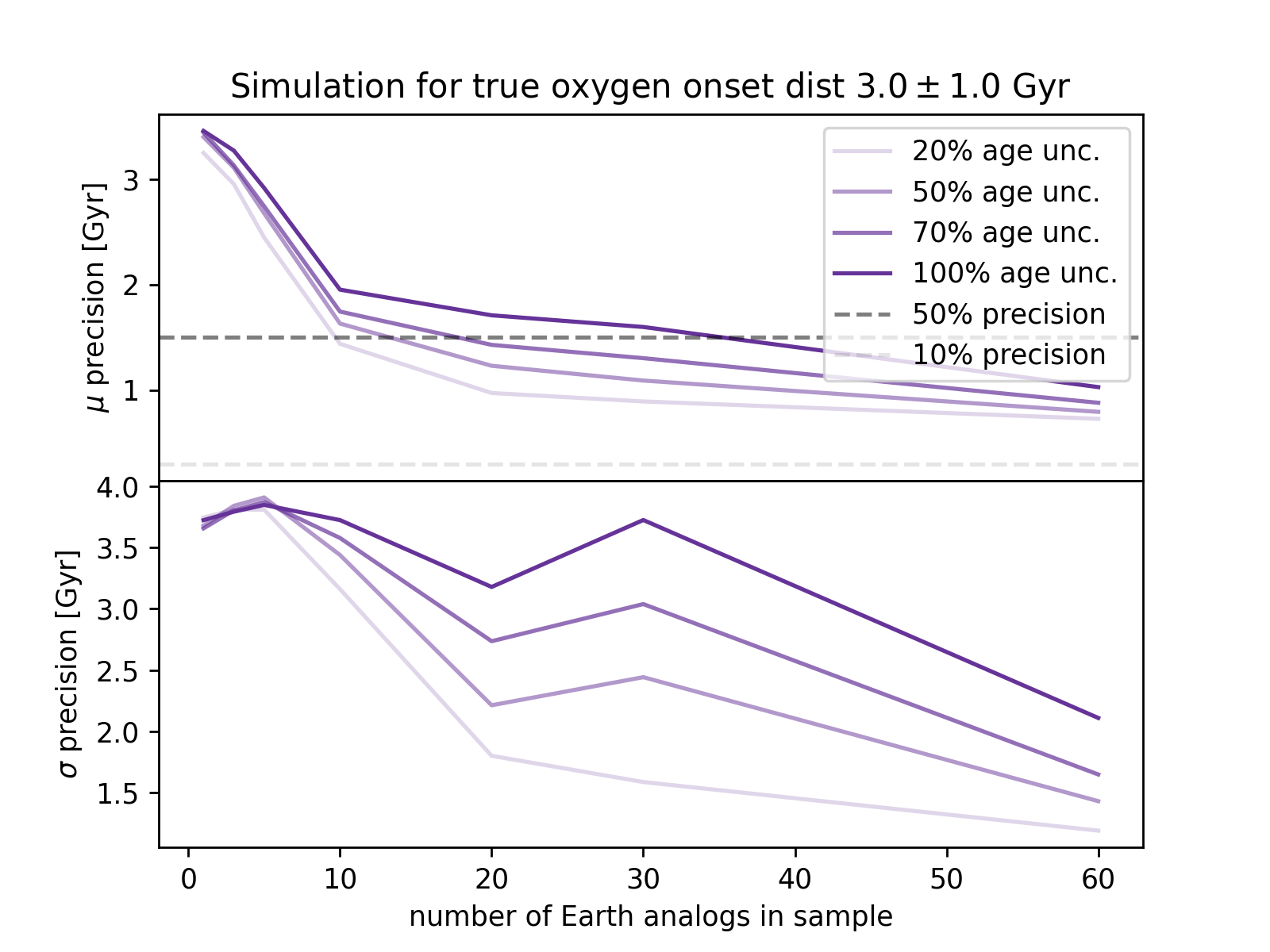}
    \caption{Same as Figure \ref{fig:2gy}, but recovering a different underlying ozone emergence time distribution. Takeaway: for a true onset time distribution with a larger mean value, age precision becomes more important. The true uncertainty of the population, $\sigma_{\rm pop}$, affects the shapes of the precision contours; i.e. for a smaller value of  $\sigma_{\rm pop}$, better precision can be obtained with fewer Earth analogs and worse age precision.
}
    \label{fig:3gy}
\end{figure}

\begin{figure}
    \centering
    \includegraphics[width=\textwidth]{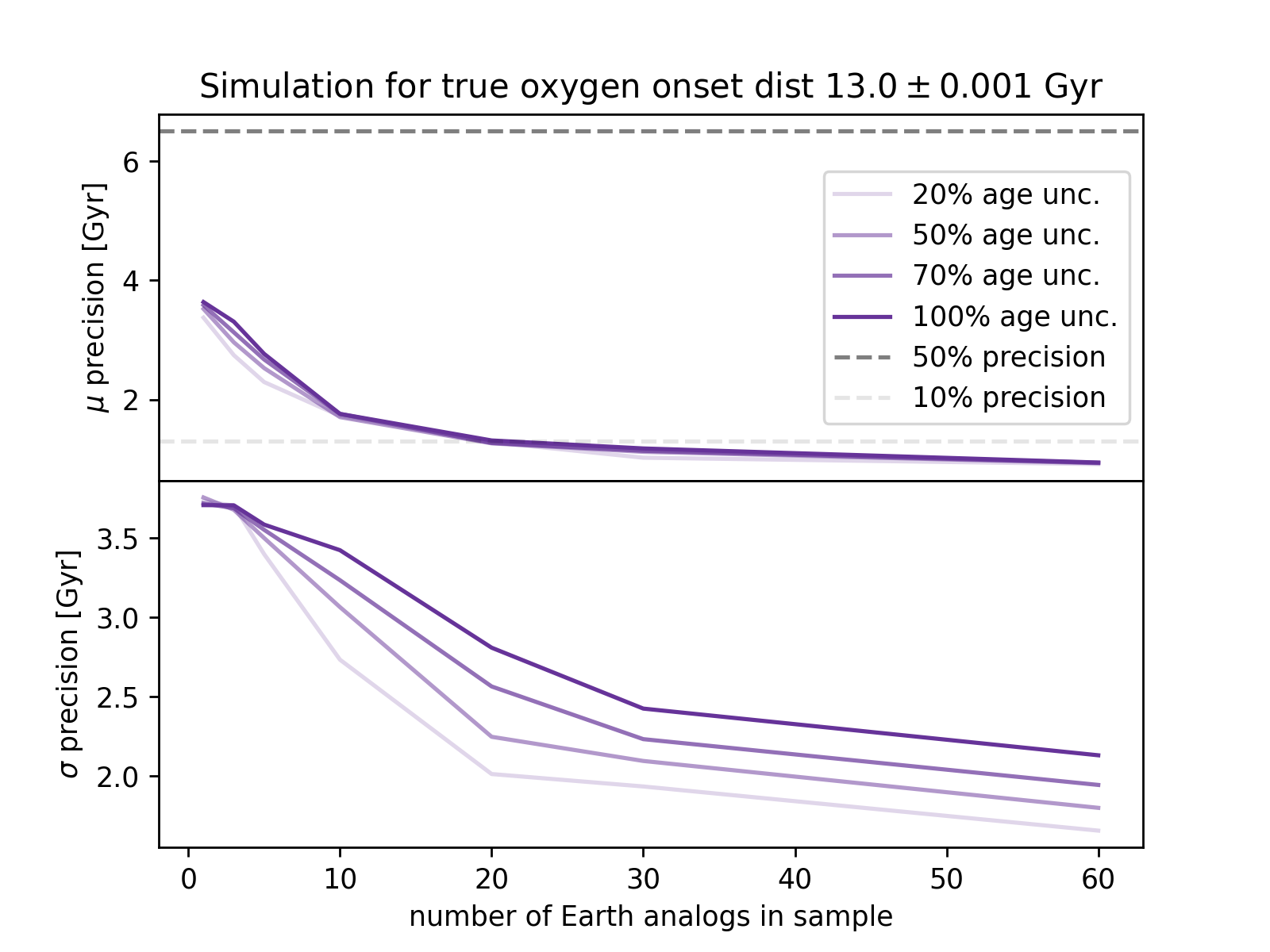}
    \caption{Same as Figure \ref{fig:2gy}, but recovering an underlying distribution with mean 13Gyr and uncertainty 0.001 Gyr, approximating the scenario that ozone never emerges on Earth analogs. Takeaway: with a population of 25 Earth analogs with no ozone detections, we can say with 10\% precision (10$\sigma$ certainty) that ozone does not emerge in the population (in other words, that the emergence time is the same as the age of the universe).
}
    \label{fig:13gy}
\end{figure}

We performed simulations for different combinations of $\sigma_{\rm pop}$ and $\mu_{\rm pop}$, stellar age uncertainty, and sample size. We display three examples of these experiments in Figures \ref{fig:2gy}, \ref{fig:3gy}, and \ref{fig:13gy}. The first two figures assume different ozone onset time distributions, and demonstrate the effect of changing stellar age uncertainty and sample size, while the third figure demonstrates parameter recovery in the case when ozone \textit{never} emerges in the population of Earth analogs. Overall, sample size impacts the precision with which we can recover the ozone onset time more than stellar age uncertainty. The recovered precision on $\sigma_{\rm pop}$ and $\mu_{\rm pop}$ is approximately independent of the actual values of these parameters, and, as expected, improves as the  sample size increases. Even with sample sizes of $\sim$100 Earth analogs (significantly beyond HWO's current stated goal of imaging $\gtrsim$25 Earth analogs), $\sigma_{\rm pop}$ is not recovered to less than 100\% precision in either of the experiments shown. However, 50\% precision (a 2$\sigma$ measurement) for $\mu_{\rm pop}$ is feasible for the scenarios shown in Figures \ref{fig:2gy} and \ref{fig:3gy} with a sample size of 30 Earth analogs. In addition, Figure \ref{fig:13gy} shows that with a sample of 30 Earth analogs \textit{without any} ozone detections, we can say with 10$\sigma$ confidence that ozone never emerges in the population (strictly, we constrain the emergence time to be the age of the universe with 10$\sigma$ precision).

\section{Discussion: Ingredients in Stellar Age Measurements and Sample Size}
\label{sec:discuss}

To constrain the ozone onset time distribution, we need a sample of Earth analog planets with known ages. In this section, we briefly discuss these requirements in order to make our simulations in the previous sections more concrete. However, we emphasize that we do not intend for the simulations we have performed here to be a ``final result'' to be interpreted at face value as requirements for HWO. We hope that future work will use the framework we have developed here to perform more realistic end-to-end simulations of HWO's capability to constrain population-level constraints about the atmospheric evolution of Earth analogs.

\subsection{Stellar Ages from Precursor Observations}

\begin{figure}
    \centering
    \includegraphics[width=0.5\textwidth]{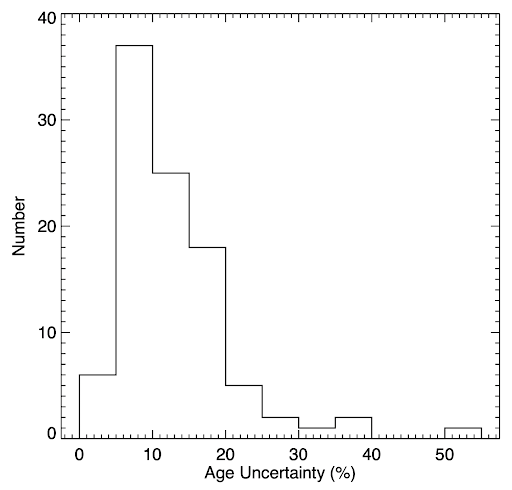}
    \caption{Distribution of age precisions of asteroseismically-derives ages of solar-type main sequence stars typically reach 10 to 20\% precision, based on an analysis of 97 stars (with masses between 0.75 and 1.5 M$_{\odot}$) with Kepler data performed by \cite{Bellinger:2019a}. Modern asteroseismology methods are able to reach 20\% precision, with high-cadence precursor observations (in this case, Kepler data).
}
    \label{fig:ages}
\end{figure}

In order to identify trends in exoplanet atmospheric demographics as a function of age, we must know the ages of the planets we are observing. Oxygenation events in Earth history are constrained to have occurred on relatively short timescales (250-500 Myr), beginning $\sim$2 Gyr after the formation of Earth, and with the rise to modern levels another approximately 2 Gyr later\cite{Lyons:2014a}. Planet formation is thought to happen quickly after star formation, an inference that is supported by the typical observed time of protoplanetary disk dispersal ($\sim$3Myr\cite{Richert:2018a}), as well as inferences about the isotopic age of the Earth (e.g. \cite{Manhes:1980a}), so we can safely assume a planet's age is approximately the same as its star's (not including post-formation giant impacts, which may last up to hundreds of millions of years \cite{Raymond:2022a}). However, a star’s age can only be inferred through the impact of stellar evolution on observable parameters, such as seismic modes, Li and Be abundances, and the decline of rotation and magnetic-related phenomena (e.g. \cite{Bouma:2024a,Bellinger:2019a,Choi:2016a,Stanford-Moore:2020a,Jeffries:2023a}). In a typical case, uncertainty on the asteroseismic age of a solar-type, core hydrogen-burning star with high-cadence data (e.g. from Kepler) is about 10-20\% (see Figure \ref{fig:ages}), while the isochrone-derived age of the same star may be 100\% or more (e.g. \cite{Saffe:2005a}).  

Precursor high-cadence photometric observations of the HWO target list, intended to resolve asteroseismic modes and derive precise ages, will likely greatly improve our ability to measure $\sigma_{\rm pop}$ (see Figures \ref{fig:2gy} and \ref{fig:3gy}). This is one of the main goals of the PLATO mission: to provide asteroseismic age estimates for 20,000 nearby main sequence stars, which should encompass HWO targets fainter than 4th magnitude, the magnitude limit of PLATO \cite{Rauer:2014a}. However, the current PLATO observation concept\footnote{https://www.cosmos.esa.int/web/plato/observation-concept} would only cover 10-50\% of the sky, motivating targeted asteroseismic follow-up for HWO targets not covered by the main PLATO mission. Currently, only $\sim$10\% of HWO Tier 1 \cite{Mamajek:2024a}\footnote{Please note that this is a provisional list of stars intended to guide precursor science efforts, and not a final target list for HWO} targets have published asteroseismic ages, but TESS 20-second cadence observations and EPRV campaigns, combined, have the capability to deliver asteroseismic ages for $\sim$90\% of these targets (Huber, private comm.)

\subsection{Earth Analog Sample}

The sample of Earth analogs that can be searched for atmospheric ozone will be identified through HWO observations of the input target list, plus follow-up observations of planet candidates to confirm each is indeed an Earth analog. Here, we define an Earth analog as a planet with a size similar to Earth (0.8 - 1.4 R$_{\oplus}$)\footnote{Where, following \cite{LUVOIR} we set the upper limit as a conservative boundary between rocky and gaseous planets recovered by \cite{Rogers:2015a}, and the lower limit as the radius below which planets at 1au are less likely to retain atmospheres \cite{Zahnle:2017a}}, with an orbit that places it fully within that star’s habitable zone (i.e. we do not include planets with significant orbital eccentricity that leave and reenter the habitable zone over their orbits). Future work (Abbas et al, in prep) will explore the number of HWO observation epochs needed to determine whether a newly-discovered Earth-like planet remains in the habitable zone over its orbit.

For $\eta_{\oplus}=25$\%\footnote{Where we assume a ``true'' Earth analog has an eccentricity small enough not to leave the habitable zone over its orbit.}, this would require an initial target list of 160 stars (for which HWO is sensitive to orbiting Earth analogs) in order to detect and confirm 40 Earth analogs.\footnote{As a point of comparison, \cite{Mamajek:2024a} defined a list of 164 stars across three tiers, where a tier A star is the most ideal for HWO observations, while a tier C star is least ideal but still possible.} If $\eta_{\oplus}$ is lower, we would consequently constrain $\sigma_{\rm pop}$ and $\mu_{\rm pop}$ to lower precision, or require a HWO architecture sensitive to more Earth analogs in order to place the same constraints.

The discovery and confirmation of Earth analogs will require the unique capabilities of HWO, though precursor radial velocity measurements, together with  may supplement HWO observations by discovering and/or placing dynamical mass constraints on some targets \cite{Harada:2024a}.

\section{Conclusion}
\label{sec:conclude}

In this study, we simulated populations of binary ozone/no ozone measurements based on assumed true population-level distributions of when ozone emerges on Earth analogs. We then evaluated our ability to recover the underlying parameters as a function of stellar age uncertainty and sample size. Our main conclusion is that, overall, sample size of Earth analogs for which HWO has the capability to robustly detect ozone is more important than stellar age uncertainties in constraining the parameters of the population, particularly the population mean. 20\% stellar age uncertainties, expected to be feasible from targeted long-duration ($\sim$years) asteroseismology measurements, are both a realistic expectation for the target stars in the Earth analog sample, and sufficient to constrain the population mean to at least 50\% precision for all three underlying true scenarios shown.


One important limitation of this study, with potential consequences that we will defer to future work, is that our simulations do not account for the possibility that ozone emerges on Earth analogs and then goes away. If this occurs, our results would be biased. Addressing this limitation should be straightforward; rather than simulating (and recovering) a distribution where ozone emerges and then persists, we could add an ``ozone lifetime'' parameter to our simulations. In addition, we do not test our ability to distinguish between ozone onset time distributions that are non-Gaussian; for example, future work could evaluate our ability to recover a bimodal distribution of onset times, corresponding to the physical scenario that some Earth analogs develop life, while others never do. 

In addition, our study purposefully presents results in terms of ``number of Earth analogs in sample,'' for simplicity. Future work should more augment this work by taking into account varying values of $\eta_{\oplus}$, atmospheric retrieval analyses, and other statistical formulations of the characteristic distribution of ozone onset times (e.g. \cite{Bixel:2021a}) in order to create an end-to-end simulation of HWO's ability to constrain this distribution.

Overall, our study emphasizes the importance of UV capability for HWO. Studies like \cite{Latouf:2024a} and \cite{Damiano:2023a} highlight the need for UV coverage for atmospheric oxygen constraint, and our study takes a step toward the conclusion that with this technological capability, we may begin to answer questions \textit{beyond} the binary ``Are we alone?'' At the conclusion of the HWO main mission, we may also be in a position to address exciting questions like ``Why are we alone?'' or ``How alone are we?''

\subsection* {Acknowledgments}
This analysis was performed as part of the HWO Demographics \& Architectures Working Group steering committee. S.B. and E.L.N. would like to thank the members of the HWO Living Worlds Working Group for very helpful conversations. S.B. gratefully acknowledges question 127086 on MathOverflow (answered by user Did) for an illuminating integral evaluation, and Asif Abbas, Dan Huber, and Jason Wang for helpful discussion. E.R.N. thanks Sarah Slotznik and Brenhin Keller for helpful conversations about the GOE. All of the authors thank the referees for thorough and constructive feedback that greatly improved this paper.

E.L.N was supported in part by NASA grants 80NSSC21K0958 and 21-ADAP21-0130. T.D. acknowledges support from the McDonnell Center for the Space Sciences at Washington University in St. Louis. C.K.H.\ acknowledges support from the National Science Foundation (NSF) Graduate Research Fellowship Program (GRFP) under Grant No.~DGE 2146752. M.R. acknowledges support by the Heising-Simons Foundation through Grants $\#$2021-2802 and $\#$2023-4478.

\subsection*{Disclosures}
The authors declare there are no financial interests, commercial affiliations, or other potential conflicts of interest that have influenced the objectivity of this research or the writing of this paper.

\subsection*{Code and Data Availability Statements}
 All code needed to reproduce the analysis in this paper is publicly available at \url{https://github.com/sblunt/HWO\_calculations}. 


\bibliography{article}   
\bibliographystyle{spiejour}   

\listoffigures
\listoftables

\end{spacing}
\end{document}